\documentclass[twocolumn, tighten, times]{aastex631}

\usepackage{xcolor}
\usepackage{soul}
\usepackage{amsmath}

\shorttitle{PKS~1717+177: blazar-UHECR correlation}
\shortauthors{Das et al.}
\graphicspath{{./}{figures/}}

\begin{document}

\title{Cosmic Clues from Amaterasu: Blazar-Driven Ultrahigh-Energy Cosmic Rays?}

\author[0000-0001-5796-225X]{Saikat Das}
\affiliation{Department of Physics, University of Florida, Gainesville, FL 32611, USA}
\email{saikatdas@ufl.edu}

\author[0009-0007-0381-8069]{Srijita Hazra}
\affiliation{Astronomy \& Astrophysics Group, Raman Research Institute, Sadashivanagar, Bangalore 560080, Karnataka, India}

\author[0000-0002-1188-7503]{Nayantara Gupta}
\affiliation{Astronomy \& Astrophysics Group, Raman Research Institute, Sadashivanagar, Bangalore 560080, Karnataka, India}



\begin{abstract}
The detection of the Amaterasu event of energy 244 EeV by the Telescope Array, one of the most energetic ultrahigh-energy cosmic rays (UHECRs; $E\gtrsim0.1$ EeV) observed to date, invites scrutiny of its potential source. We investigate whether the nearby blazar PKS~1717+177 at redshift $z=0.137$, located within $2.5^\circ$ of the reconstructed arrival direction, could explain the event under a proton-primary hypothesis. Using a one-zone jet model, we fit the multiwavelength spectral energy distribution of the source, incorporating both leptonic and hadronic cascade emissions from photohadronic interactions inside the jet. Our model supports a cosmic-ray origin of the very-high-energy ($\varepsilon_\gamma\gtrsim 100$ GeV) $\gamma$-ray flux and predicts a subdominant neutrino flux, one order of magnitude lower than from TXS~0506+056. Under Lorentz invariance violation, UHECRs escaping the blazar jet above a specific energy can propagate unattenuated over hundreds of Mpc due to an increase in energy loss length for certain parameter choices.
In such a scenario, the Amaterasu event can have a plausible origin from this blazar. Our analysis indicates negligible deflection in the Galactic magnetic field, implying a strong extragalactic magnetic field is required. Our findings provide a compelling multimessenger framework linking UHECRs, $\gamma$-rays, and neutrinos and motivate targeted searches by current and future high-energy neutrino telescopes during increased $\gamma$-ray or X-ray activity of this blazar.
\end{abstract}



\section{Introduction} \label{sec:intro}
The origin of ultrahigh-energy cosmic rays (UHECRs; $E\gtrsim 10^{17}$ eV) is an unsolved enigma in astrophysics \citep[see][for review]{Kotera_2011, Anchordoqui:2018qom}. Unlike $\gamma$-rays and neutrinos, their sources are obscured by deflections in cosmic magnetic fields \citep{Sigl:2004yk, Globus:2007bi}. 
Observed features in their energy spectrum suggest a distinct extragalactic source population. However, the exact transition energy between Galactic and extragalactic cosmic rays is still unknown \citep[e.g.,][]{Aloisio:2012ba}. 
The leading experiments observing these particles are the Pierre Auger Observatory in Argentina \citep{PierreAuger:2015eyc} and the Telescope Array (TA) in Utah, USA \citep{TelescopeArray:2015dcv}.

The detection of the ``Oh-My-God'' particle with an energy of $\approx320$ EeV by the Fly’s Eye experiment is the most energetic event recorded using the fluorescence technique, capturing the air shower due to the interaction with Earth's atmosphere \citep{HIRES:1994ijd}. However, the advent of water Cherenkov surface detector arrays has boosted the duty cycle of UHECR detection. The `Amaterasu' event observed by TA on 2021 May 27, named after the Shinto sun goddess for its brilliance, is one of the most energetic cosmic-ray events detected by the surface detector. The estimated reconstructed energy of $244 \pm 29 \mathrm{\ (stat)}^{+51}_{-76} \mathrm{\ (sys)}$ EeV and direction (RA, decl.) = ($255.9^\circ\pm 0.6^\circ$, $16.1^\circ\pm 0.5^\circ$) places it among the most energetic particles ever observed \citep{TelescopeArray:2023sbd}.

The detection of such an extreme-energy event raises important questions regarding the nature and location of its source. 
The Greisen-Zatsepin-Kuz'min suppression \citep{Greisen_1966, Zatsepin_1966} limits the observed UHECR energy due to photopion production off the cosmic microwave background (CMB). The energy-loss mean free path of protons at this energy is $\mathcal{O}\sim10$~Mpc \citep[see, e.g.,][]{Dermer:2008cy}. UHECR composition studies by TA suggest light nuclei at $\gtrsim 10^{19}$ eV, with significant uncertainties \citep{TA_2010}. A combined fit of the spectrum and composition
by Auger suggests progressively heavier nuclei at $\gtrsim 10^{18.2}$ eV \citep{PierreAuger:2016use}. The maximum proton fraction in the highest-energy bin of the UHECR spectrum can go up to $\sim10\%-15\%$ 
\citep{Muzio:2019leu, Das:2020nvx, Ehlert:2023btz}. If the Amaterasu event originated within the local GZK horizon for protons, 
a source correlation is expected owing to the high magnetic rigidity. However, Amaterasu’s direction points to the Local Void, a region of the sky with a relatively small number of galaxies \citep{Tully:2007ue}.

Ultraheavy UHECRs \citep{Zhang:2024sjp}, binary neutron star mergers \citep{Farrar:2024zsm}, superheavy dark matter decay \citep{Murase:2025uwv, Sarmah:2024ffy}, the scattering of ultrahigh-energy (UHE) neutrinos by the cosmic neutrino background through Z-boson resonance \citep{Fargion:2024ujt}, starburst galaxy \citep{Bourriche:2024bbe}, and a transient event in an unresolved Galaxy \citep{Unger:2023hnu} have been proposed for the astrophysical explanation of the Amaterasu event. UHECRs serve as a sensitive probe for testing Lorentz Invariance Violation \citep[LIV;][]{PierreAuger:2021tog}, and such a possibility is also explored for this event \citep{Lang:2024jmc}. 

The reconstructed arrival direction of the Amaterasu event lies within $2.5^\circ$ of the blazar PKS~1717+177 at redshift $z = 0.137$, assuming a proton primary. It is a nearby active galactic nucleus (AGN) identified in the $\gamma$-ray source catalog \citep{Fermi-LAT:2019yla}.
This flaring $\gamma$-ray source has been proposed 
as a neutrino-emitting AGN in point source stacking analysis of IceCube data \citep{Britzen:2024egp}. 
AGNs have long been considered as sources of high-energy cosmic rays and neutrinos \citep[see, e.g.,][]{1979ApJ...232..106E, 1981MNRAS.194....3B, Sikora_1987, 1991PhRvL..66.2697S, 1992A&A...260L...1M, Szabo:1994qx, Atoyan:2001ey, Murase:2014foa, Das:2020hev, Das:2021cdf, Murase:2022feu}. The detection of a high-energy neutrino by the IceCube Observatory in spatial and temporal coincidence with the flaring $\gamma$-ray blazar TXS~0506+056 corroborates cosmic-ray acceleration in jetted AGNs and its imprints on multiwavelength spectrum \citep[e.g.,][]{IceCube:2018cha, IceCube:2018dnn, MAGIC:2018sak, Keivani:2018rnh}.

In this work, we show that UHE protons of energy $\gtrsim 0.1$ EeV can be injected from the jet of this blazar.
They undergo deflection in cosmic magnetic fields without suffering interactions beyond 10 EeV for a suitable choice of the LIV coefficient \citep{PierreAuger:2021tog}. However, the LIV effects are insignificant below tens of EeV energies. Hence, a hadronic signature in the $\gamma$-ray flux should be observed due to interactions inside the jet if this source accelerates cosmic rays.  

A fit to the blazar spectrum using a leptohadronic model is obtained, including radiation from the electromagnetic cascade of secondary electrons. 
Our analysis reveals that the fit to very-high-energy (VHE; $E\gtrsim0.1$ TeV) $\gamma$-ray flux is improved compared to a purely leptonic model. The latter suffers from the Klein-Nishina effect, resulting in inefficient inverse-Compton scattering at the VHE regime. The estimated neutrino event rate is beyond the sensitivity of current detectors.
The required kinetic power in protons constrains the rigidity cutoff required in the UHECR spectrum. We also analyze the deflection in cosmic magnetic fields, thus constraining the rms field strength of the extragalactic medium. 

We present our analysis and results obtained in Sec.~\ref{sec:results} and examine the implications for UHECR sources in Sec.~\ref{sec:discussions}. We draw our conclusions in Sec.~\ref{sec:conclusions}.

\section{Model considerations \label{sec:model}}

The broadband spectral energy distribution (SED) of the source is obtained from the
archival data retrieved from the SSDC SED builder \citep{Stratta_2011}.
The radio data were obtained from CRATES \citep{2007ApJS..171...61H}, FIRST \citep{1994ASPC...61..165B}, NIEPPOCAT \citep{2007AJ....133.1947N}, PKSCAT90 \citep{1992BICDS..41...47W}, and PLANCK \citep{2014A&A...571A..28P}.
The infrared (IR) data were collected from the WISE catalog \citep{2010AJ....140.1868W}. The data for optical-UV wave bands were collected from the UV and Optical Telescope (UVOT) of the Swift Observatory \citep{2012A&A...541A.160G} and the GALEX \citep{2011Ap&SS.335..161B} catalog. The X-ray data were found from Swift-XRT \citep{Britzen:2024egp} and MAXI \citep{Kawamuro:2018eky}. 
The Fermi 4FGL-DR4 \citep{2023arXiv230712546B}, along with earlier data releases,  
provides the $\gamma$-ray spectrum of the source.


We model the emission region of PKS~1717+177 as a spherical blob of radius $R'$ embedded in the jet, consisting of a relativistic plasma of electrons and protons moving through a uniform magnetic field $B^\prime$ in the comoving jet frame. The jet has a bulk Lorentz factor $\Gamma$, and the Doppler factor is $\delta_D = [\Gamma(1 - \beta \cos\theta)]^{-1}$, where $\beta c$ is the plasma velocity and $\theta$ the viewing angle. For $\theta \lesssim 1/\Gamma$, we approximate $\delta_D \approx \Gamma$.
Electrons are injected with a power-law distribution,
\begin{align}
    Q'_e(\gamma'_e) = A_e \left( {\gamma'_e}/{\gamma_0} \right)^{-\alpha} \quad \text{for } \gamma'_{e, \min} < \gamma'_e < \gamma'_{e, \max},
\end{align}
to reproduce the observed broadband SED. Here, $A_e$ depends on the injected electron luminosity, and $\gamma_0 m_e c^2 = 500$~MeV is a reference energy. A quasi-steady state is achieved when the injection is balanced by radiative cooling and/or escape, yielding $N'_e(\gamma'_e) = Q'_e(\gamma'_e)\, t'_e$
with $t'_e = \min(t'_{\rm cool}, t'_{\rm esc})$, and $t'_{\rm esc} \simeq t'_{\rm dyn} = 2R'/c$. The cooling timescale is
\begin{align}
    t'_{\rm cool} = {3 m_e c}/{4 \sigma_T \gamma'_e (u'_B + \kappa_{\rm KN} u'_{\rm ph})},
\end{align}
where $u'_B = B^{\prime 2} / 8\pi$, and $u'_{\rm ph}$ is the soft photon energy density. $\kappa_{\rm KN}$ accounts for Klein-Nishina suppression of inverse-Compton emission.
We solve the  transport equation using the open-source code \textsc{GAMERA} \citep{Hahn:2015hhw, Hahn_2022} to find the steady-state solution to
\begin{align}
    \frac{\partial N'_e}{\partial t} = Q'_e(\gamma'_e,t') - \frac{\partial}{\partial \gamma'_e}(b N'_e) - \frac{N'_e}{t_{\rm esc}},
\end{align}
where $b(\gamma'_e, t')$ is the energy loss rate.
The steady-state electron spectrum produces synchrotron (SYN) and synchrotron self-Compton (SSC) emission. We include external inverse-Compton (EC) scattering of a blackbody photon field with temperature $T'$ and comoving energy density $u'_{\rm ext} = (4/3)\Gamma^2 u_{\rm ext}$, where $u_{\rm ext} = \eta_{\rm ext} L_{\rm disk} / 4\pi R_{\rm ext}^2 c$ is the energy density in the AGN frame, and $\eta_{\rm ext}$ is the fraction of the disk luminosity \citep{Barnacka:2013oxa}. The emission region is located at a distance $R_{\rm ext}$ along the jet axis. The external photons enter the jet and are Doppler-boosted in the comoving frame. The EC model provides a good fit to the entire broadband spectrum,  
suggesting that the source could be a masquerading BL Lac, i.e., an intrinsically flat-spectrum radio quasar with a hidden broad-line region (BLR) and a standard accretion disk, as proposed for TXS~0506+056 \citep{Padovani:2019xcv}.

We follow the numerical method described in \citet{Das:2022nyp} and \citet{Prince:2023bhj} to model hadronic emission. The proton injection at the source is given by a power-law spectrum with a rigidity-dependent cutoff,
\begin{equation}
    \dfrac{dN}{dE} = \begin{cases}
        A_p E_p^{-\alpha} & (E_p<ZR_{\rm max})\\
        A_p E_p^{-\alpha}\exp\bigg(1-\frac{E_p}{ZR_{\rm max}}\bigg) & (E_p\geq ZR_{\rm max})
    \end{cases}
    \label{eqn:injection}
\end{equation}
For protons interacting in the blazar jet, $E_p\ll R_{\rm max}$. Their dominant energy losses are through photopion production ($p\gamma \rightarrow p+\pi^0$ or $n+\pi^+$) and the Bethe–Heitler (BH) process ($p\gamma \rightarrow p+e^+e^-$), with target photons originating from leptonic emission and external photon field. Charged pions decay to produce neutrinos. The interaction timescale is given by an integral over the photon distribution, cross section, and inelasticity as functions of photon energy in the proton rest frame \citep{Stecker:1968uc, Chodorowski_92, Mucke_00}. Below a few PeV, the proton interaction timescale is longer than the dynamical timescale $t'_{\rm dyn}$. To achieve a significant $\pi^0$-decay $\gamma$-ray flux, we model the escape rate such that it is smaller than the $p\gamma$ interaction rate and hence can be neglected. The normalization $A_p$ is set by the required kinetic power in protons. $\gamma$-ray and $e^-$ spectrum from pion decay and $e^\pm$ pairs from the BH process are calculated using the parameterization by \citet{Kelner:2008ke}, weighting the input proton spectrum by the respective interaction rate. For example, to calculate the electron spectrum from BH interactions, protons are injected with a spectrum $N'_p(\gamma'_p) \times R_{\rm BH}/R_{\rm tot}$, where $R_{\rm BH}$ and $R_{\rm tot}$ are the BH and total (BH + photopion) interaction rates, respectively.

High-energy $\gamma$-rays undergo $\gamma\gamma\rightarrow e^\pm$ absorption with soft photons and the external radiation field. The escaping TeV spectrum is thus suppressed as
$Q^\prime_{\gamma, \rm esc}(\epsilon^\prime_\gamma) = Q^\prime_{\gamma, \pi}(\epsilon^\prime_\gamma)({1 - e^{-\tau_{\gamma\gamma}}})/{\tau_{\gamma\gamma}}$
where $\tau_{\gamma\gamma}$ is the optical depth calculated following \citet{Gould_1967}, integrating over the target photon distribution and pair-production cross section.
The high-energy $\gamma\gamma$ pair production spectrum ($Q'_{e,\gamma\gamma}$) is solved numerically using the expression in \cite{Boettcher_1997}. Together with the $e^-$-s from charged pion decay ($Q'_{e,\pi}$) and BH process ($Q'_{e,\rm BH}$), they can initiate cascade radiation from the jet. We compute the steady-state spectrum $N'_{e,s}$ of these electrons using the semianalytical approach of \citet{Boettcher:2013wxa}, incorporating $Q'_{e, \rm BH}$ in the source term and using the same escape timescale as for primary electrons. In synchrotron-dominated cascades, the resulting photon spectrum is obtained by integrating over the electron distribution weighted by the SYN kernel $
Q'_s(\epsilon'_s) = A_0 \epsilon'^{-3/2}_s \int_1^\infty N'_{e,s}(\gamma'_e) \gamma'^{-2/3}_e e^{-\epsilon'_s / b\gamma'^2_e} \, d\gamma'_e$
with 
$A_0 =c\sigma_TB'^2/[6\pi m_e c^2 \Gamma(4/3)b^{4/3}]$ being a normalization constant, 
%
%
where $b=B'/B_{\rm crit}$ and $B_{\rm crit} = 4.4\times 10^{13}$ G. 

\section{Results\label{sec:results}}

\subsection{Multimessenger emission}

\begin{figure}
    \centering
    \includegraphics[width=0.49\textwidth]{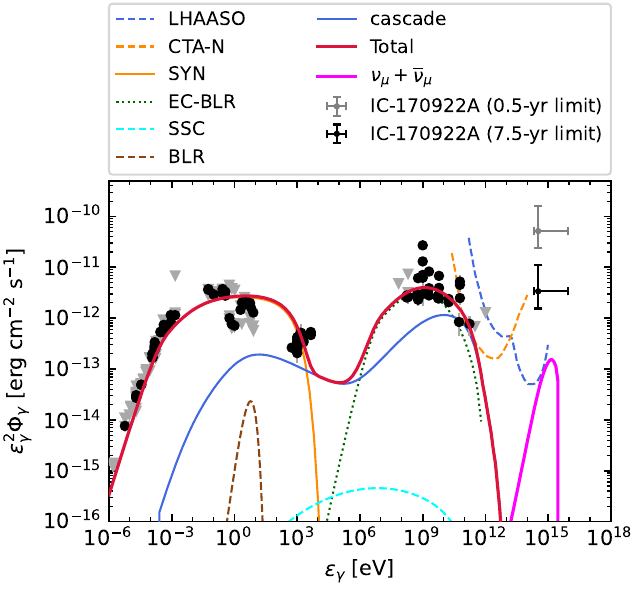}
    \caption{Multiwavelength SED of PKS~1717+177 showing the synchrotron (orange solid), external Compton (green dotted), the external photon field (brown dashed), the hadronic cascade (blue solid), and the $\nu_\mu+\overline{\nu}_\mu$ spectrum (magenta) in the observer frame. Black data points show the spectral data from radio to VHE $\gamma$-rays, and upper limits are shown in gray. The black and gray error bars correspond to the neutrino flux from TXS 0506+056, which is shown for reference, assuming a 7.5-yr and 0.5-yr limit. The SSC spectrum shown in cyan color is negligible in the EC model. The orange dashed and blue dashed lines show the sensitivity of CTA-North and LHAASO detectors, respectively.}
    \label{fig:mwl_sed}
\end{figure}

\begin{table*}[t]
\centering
\caption{\label{tab:mwl_fit}Leptohadronic model parameters for the fit to multiwavelength SED. Electron and proton luminosities are in the AGN rest frame.}
\renewcommand{\arraystretch}{1.2}
\resizebox{\textwidth}{!}{%
\begin{tabular}{ccccccccccccc}
\hline
$\delta_D$ & $B'$ & $R'$ & $u'_{\rm ext}$ & $T'$ & $\alpha$ & $E_0$ & $\gamma'_{\rm e,min}$ & $\gamma'_{\rm e,max}$ & $L_e$ & $\gamma'_{\rm p,min}$ & $\gamma'_{\rm p,max}$ & $L_p$ \\
-- & [G] & [cm] & [erg cm$^{-3}$] & [K] & -- & [MeV] & -- & -- & [erg s$^{-1}$] & -- & -- & [erg s$^{-1}$] \\
\hline
25 & 1.0 & $5\times 10^{16}$ & 0.1 & $5\times10^5$ & 2.0 & 500 & 20 & $5\times10^4$ & $4.1\times10^{42}$ & 10 & $4.27\times10^6$ & $1.3\times10^{45}$ \\
\hline
\end{tabular}
}
\end{table*}

All model parameters were varied freely except the injection spectral index. A log-parabola injection spectrum was tested, but a simple power-law spectrum with $\alpha=2$ yields a satisfactory fit. Parameter degeneracy was assessed by scanning over a broad range of values. Given the multiple flux values at energies in the high-energy peak, we focused on the lower flux values to model the quiescent state. Earlier SED modeling of the source using older Fermi-Large Area Telescope (LAT) data \citep{2010MNRAS.401.1570T} suggests a hard LAT spectrum up to the highest energies, requiring a rather large Doppler factor, $\delta>50$. We found that with improved spectral coverage across the radio to $\gamma$-ray band, the SSC model struggles to fit the entire dataset consistently.

Although a leptonic inverse-Compton model remains feasible, we adopt a leptohadronic scenario motivated by the observed sub-TeV $\gamma$-ray emission. Moreover, including a hadronic component enables simultaneous reproduction of the highest-energy $\gamma$-ray and soft X-ray data through secondary electromagnetic cascades, which is difficult to achieve with leptonic emission alone. Here, we obtain $\delta_D = 25$, consistent with the SED fit up to the VHE regime.

We present the obtained best-fit parameter values in Table~\ref{tab:mwl_fit}, considering a steady-state spectrum. The SED fit is shown in Fig.~\ref{fig:mwl_sed}, and the various components from leptonic and hadronic emission are labeled. 
The radius of the emission region $R^\prime$, maximum electron Lorentz factor $\gamma^\prime_{\rm e, max}$, and the magnetic field $B^\prime$ are determined by fitting the optical and radio data with SYN emission (orange solid line). The latter peaks in the optical band, and a power-law injection spectrum of electrons provides a good fit to the data.
The SYN radiation of muons and pions produced in hadronic interactions is not significant for  $B^\prime = 1$ G. 
The high-energy peak constrains the temperature and energy density of the external photon field (brown dashed line), likely originating in the BLR region. They act as the most important target photon field for EC emission (green dotted line) as well as the $p\gamma$ interaction of cosmic rays. This photon field also absorbs high-energy pion-decay $\gamma$-rays through $\gamma\gamma$ pair production.

For a typical disk luminosity of $10^{46}$ erg s$^{-1}$ and considering the scattered disk emission to be a fraction $\eta_{\rm disk}\sim 0.01$ of the disk photon energy density, $R_{\rm ext}$ comes out to be $\sim0.5$ pc. The secondary electromagnetic radiation (blue solid line) shows two peaks, viz., a lower energy peak from SYN radiation of secondary electrons, dominated by BH pairs, and a high-energy spectrum from pion-decay $\gamma$-rays. The VHE $\gamma$-ray data constrain the contribution from the pion-decay cascade and hence the proton spectrum normalization, limiting the peak flux of neutrinos.
In the absence of hard X-ray data in the $10-100$ keV range, the BH component of the hadronic cascade does not provide additional constraints \citep[see, e.g.,][]{Das:2022nyp, Jiang:2025hqb}.

High-energy $\gamma$-ray absorption by the extragalactic background light (EBL), composed of optical, infrared, and ultraviolet photons, is included using the \citet{Gilmore:2011ks} model.
Observations by the Large High Altitude Air Shower Observatory \citep[LHAASO;][]{Vernetto:2016gro} and the Cherenkov Telescope Array \citep[CTA;][]{Gueta:2021vO} will further constrain the TeV $\gamma$-ray flux and hence the hadronic emission component. 
The orange dashed line in
Fig.~\ref{fig:mwl_sed} shows the differential point source sensitivity of the CTA northern array, assuming 50 hr observation time and pointing to 20$^\circ$ zenith angle. The blue dashed line shows the LHAASO 1 yr sensitivity to a Crab-like $\gamma$-ray point source. 

A cutoff in the proton spectrum beyond a maximum energy $E'_{p,\max}$ is required to explain the VHE spectrum. This also limits the peak energy of the neutrino spectrum. We assume protons above this energy escape the source, with an energy-independent escape timescale $\sim R/c$ sufficient to suppress $p\gamma$ interactions inside the jet. If the diffusion is faster than $\propto E^1$, interaction efficiency rapidly declines above tens of PeV. This is reasonable when a quasi-ballistic propagation is assumed instead of diffusive propagation inside the jet emission region. In the one-zone model,  the efficient escape of UHECRs requires a rigidity-dependent diffusion rate, for example, $D(E)\propto E^2$ at higher energies \citep{Globus:2007bi, Harari:2013pea, Muzio:2021zud}. The resulting muon neutrino ($\nu_\mu +\overline{\nu}_\mu$) flux from $p\gamma$ interactions is given by
\begin{equation}
E_\nu^2 J_\nu = ({1}/{3})  (E_\nu'^2 Q'_{\nu, p\gamma})({V' \delta_D^2 \Gamma^2}/{4\pi d_L^2}),
\end{equation}
where the factor $1/3$ accounts for neutrino oscillations and $Q'_{\nu, p\gamma}$ is the comoving-frame production rate of electron and muon neutrinos from charged pion decay.

The magenta curve in Fig.~\ref{fig:mwl_sed} shows the resulting muon neutrino spectrum obtained using the same normalization of the proton spectrum, after accounting for neutrino oscillations. The $\nu_\mu+\overline{\nu}_\mu$ event rate at IceCube in a given operation time  $\Delta T$ is calculated using the expression
\begin{equation}
    \mathcal{N}_{\nu_\mu} =\Delta T \int_{\epsilon_{\nu, \rm min}}^{\epsilon_{\nu, \rm max}} d\epsilon_\nu \dfrac{d\Phi_\nu}{d\epsilon_\nu} \langle A_{\rm eff} \rangle_\theta
\end{equation}
where $\langle A_{\rm eff} \rangle_\theta$ is the effective detector area averaged over the zenith angle bin concerned \citep{IceCube:2018ndw}. We use the effective area for event selection as a function of neutrino energy from \cite{Stettner:2019tok}. We calculate the number of muon neutrino events from this source to be $\sim0.1$ events in $10$ yr at IceCube. It is much lower than the muon neutrino flux predicted from TXS~0506+056, assuming a 0.5-yr and 7-yr flux level, also shown in the figure for comparison. 

%
%

\begin{figure}
    \centering
    \includegraphics[width=0.46\textwidth]{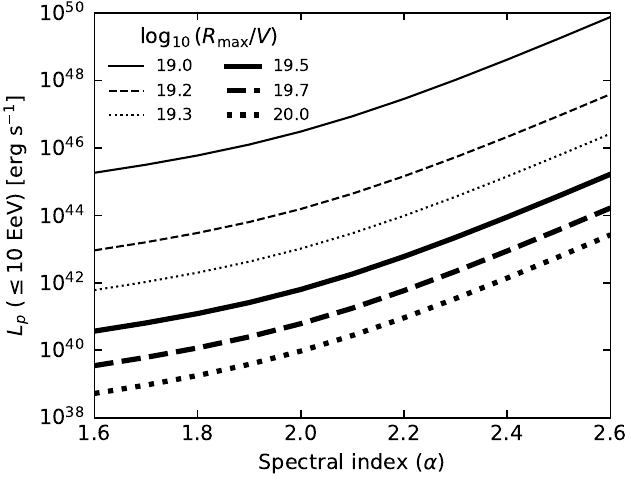}
    \caption{Proton luminosity in the 0.1 TeV to 10 EeV energy range constrained by the detection of 1 UHE proton event in the Telescope Array energy bin 148 - 340 EeV. The different curves correspond to various values of rigidity cutoff $R_{\rm max}$ in the injection spectrum.}
    \label{fig:lum_alpha}
\end{figure}

We note that the maximum proton energy required to explain the observed SED is $\simeq4$ PeV in the comoving jet frame. This corresponds to the AGN rest frame proton energy of $E_p=\Gamma E^\prime_p=0.1$ EeV. We assume protons above this energy escape the blazar jet without significant energy loss and propagate through the extragalactic medium. 



\subsection{Luminosity requirement and LIV coefficient}


LIV in the hadronic sector is defined by the modified dispersion relation $E_i^2 - p_i^2=m^2 + \delta_i^{(n)}E^{2+n}$ \citep{Coleman:1997xq}. The mean free path of UHECRs is significantly increased when $\delta_{\rm had, 0}>0$. For a choice of LIV coefficient $\delta_{\rm had,0}=10^{-21}$ in the leading order, a primary proton of energy $\mathcal{O}\sim 100$ EeV can reach Earth from a distance corresponding to this blazar without being attenuated by photopion production; suffering energy loss only due to expansion of the Universe \citep[cf. Fig.~4 in][]{PierreAuger:2021tog}. This value is well within the upper limit of $\delta_{\rm had, 0}<10^{-19}$ presented by Auger.
We constrain the value of $R_{\rm max}$ and hence the required LIV coefficient from the proton luminosity required to explain the blazar SED. 

The detection of a single event by TA in the energy range $148 - 340$ EeV (obtained by adding the statistical and systematic uncertainties in quadrature), assuming no interactions, provides the injected luminosity in this energy range for TA exposure $\mathcal{E} = 1.6\times 10^4$ km$^2$ sr yr \citep{TelescopeArray:2023sbd}. We use the injection spectrum per unit time in Eq.~\ref{eqn:injection}, emitted within a solid angle $2\pi(1-\cos\theta_{\rm jet})$ subtended by a typical jet opening angle of $\theta_{\rm jet}\approx0.1$ radians \citep{Pushkarev_2009, Finke:2018pkl}. Using the same normalization, we can calculate the kinetic power in protons for the energy range 10 TeV up to 10 EeV, beyond which LIV effects kick in.  In Fig.~\ref{fig:lum_alpha}, we show the luminosity required in protons as a function of the injection spectral index $\alpha$ for different values of $R_{\rm max}$. Noting that our multiwavelength modeling requires a proton power of $L_p = 1.3\times 10^{45}$ erg s$^{-1}$ in the blazar jet (see Table.~\ref{tab:mwl_fit}) and a spectral index of $\alpha=2$, this corresponds to a total power of $L_p (\leq 10 \text{\ EeV}) = 2\times 10^{45}$ erg s$^{-1}$. Thus, the corresponding value of $\log_{10}(R_{\rm max}/V)$ is inferred between 19 and 19.2 from Fig.~\ref{fig:lum_alpha}. The value of cutoff rigidity obtained by Auger at this LIV coefficient for a generic source population, from a combined fit to the energy spectrum and composition using Sybill2.3c hadronic interaction model \citep{Riehn:2019jet} and \cite{Gilmore:2011ks} EBL model, is $\sim 18.6$ and increases with increasing $\delta_{\rm had,0}$. 


\subsection{UHECR deflection in cosmic magnetic fields}

\begin{figure}
    \centering
    \includegraphics[trim={0 0 0 0.5cm},clip, width=0.45\textwidth]{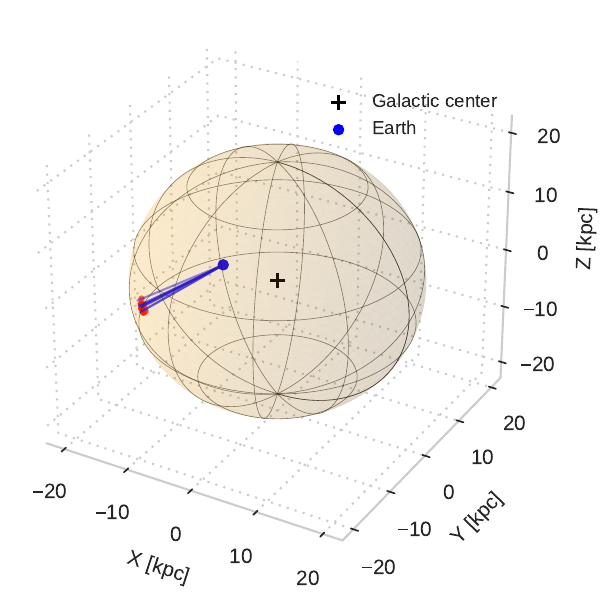}
    \caption{UHECR particles placed randomly on the surface of the Galactic horizon (red) for \cite{Unger:2023lob} magnetic field with momentum vector given by von Mises-Fisher distribution concentrated around the angular uncertainty region, centered on the best-fit direction at Earth. The black and blue dots correspond to the Galactic center and Earth. A backtracking simulation of protons from Earth to these directions shows negligible deflections in the GMF, illustrating the requirement for a high magnitude of EGMF.}
    \label{fig:gal_traj}
\end{figure}

At these extreme energies, UHECRs do not lose significant energy en route to Earth (due to LIV) but may undergo deflection in cosmic magnetic fields. The angular resolution of the TA surface detector for events with $E>10$ EeV is $1.4^\circ$ \citep{Kim:2023eul}. To account for the directional uncertainty in the observed arrival direction of the Amaterasu event, we generate sample UHECR particles with their random momentum direction drawn from a von Mises-Fisher (vMF) distribution $f(\theta)=\kappa\exp(\kappa\cos\theta)/2\sinh\kappa$. The value of the parameter $\kappa$ is chosen such that the particles are concentrated around the angular uncertainty region, centered on the best-fit direction at Earth. We show these directions intersecting the Galactic boundary as red points on the spherical surface in Fig.~\ref{fig:gal_traj}. The Galactic boundary is considered to be an outer radius of $\approx 20$ kpc from the Galactic center (black cross) in various magnetic field models \citep{Jansson:2012pc, Unger:2023lob}.

Next, we perform a backtracking simulation of cosmic rays from Earth (blue dot in Fig.~\ref{fig:gal_traj}), sampling energies given by a power-law spectrum $dN/dE\sim E^{-2}$ between the uncertainty range 148 and 340 EeV. These cosmic rays are propagated with injected directions drawn from a distinct vMF sample to check the deflection at the Galactic boundary due to the Galactic magnetic field (GMF). We use the Monte Carlo simulation framework CRPROPA 3.2 for the backtracking simulation \citep{AlvesBatista:2016vpy, AlvesBatista:2022vem}. The code allows the implementation of various magnetic field models for the propagation of UHECRs. 

We adopt the UF23 model, as introduced by \cite{Unger:2023lob}, which provides a divergence-free parameterization of the large-scale disk, toroidal halo, and poloidal (X-shaped) field components. This model is constrained by a joint fit to full-sky rotation measures and polarized synchrotron intensity maps from WMAP and Planck, using updated electron density and cosmic-ray lepton distributions. We test the 20-parameter \texttt{base} model as well as the local magnetic spur structure near the Sun (\texttt{spur}) and exponential decay (\texttt{expX}) models. We performed the backtracking simulation using 10 particles across 100 random realizations of the vMF. The resulting intersection points on the Galactic boundary align well with the initial vMF distribution for all models, indicating that deflections due to the GMF are negligible. Fig.~\ref{fig:gal_traj} shows that UHECRs at this energy travel along a straight line path in the Galaxy. The results were also found to be consistent with the JF2012 model \citep{Jansson:2012pc}. Hence, the deflection from the direction of the blazar can be attributed to the extragalactic magnetic field (EGMF).

A high intensity of the EGMF is expected for the $2.5^\circ$ deflection from the blazar. For Larmor radius $r_L$ in the magnetic field, the expected deflection angle is $\theta_{\rm dfl}= {\sqrt{2d_c\lambda_c}}/{3r_L}$, assuming small-angle scattering. We calculate the required strength of EGMF and express the deflection of a UHECR with charge $Z$ from the blazar direction in the parametric form as  \citep{Dermer:2008cy, Murase:2011yw},
\begin{align}
    \theta_{\rm dfl} \simeq 2.5^\circ\, Z 
    &\left( \frac{d_c}{590~\mathrm{Mpc}} \right)^{1/2}
    \left( \frac{\lambda_c}{500~\mathrm{kpc}} \right)^{1/2}  \nonumber \\
    &\times\left( \frac{B_{\rm EG}}{1.54~\mathrm{nG}} \right)
    \left( \frac{E_{p}}{250~\mathrm{EeV}} \right)^{-1}
\end{align}
where $d_c=590$~Mpc is the comoving distance to the source, $\lambda_c$ is the turbulence coherence length, and $B_{\rm EG}$ is the rms strength of the EGMF. We find that $B_{\rm EG}\approx 1.5$ nG is required to deflect protons by $2.5^\circ$ from the initial emission direction. The value corresponds to the magnetic field in cosmic filaments \citep{Vazza:2014jga}. The Universe is not homogeneous at these distance scales, and UHECRs may encounter matter concentrated in superclusters or filaments, separated by intergalactic voids, resulting in the deflection of proton primaries.

The Telescope Array collaboration showed backtracked arrival directions assuming two GMF models, JF2012 and PT2011 \citep{Pshirkov:2011um}. In both cases, a plausible association with blazar PKS 1717+177 emerges only for a proton primary. For heavier nuclei, such as iron, the backtracked direction in the JF2012 model shifts toward a region of the large-scale structure populated by galaxies \citep[cf. Fig. 2 in][]{TelescopeArray:2023sbd}, disfavoring the blazar hypothesis.

\section{Discussions\label{sec:discussions}}

Earlier studies on the broadband emission of this source imply a high Doppler factor $\delta_D>50$ and a low magnetic field for the SSC model. However, it is discussed that the SSC model does not provide a good fit to the LAT $\gamma$-ray spectrum under a one-zone leptonic model only \citep{2010MNRAS.401.1570T}. In our work, we provide a first analysis of this source in light of leptohadronic interactions. The nuclear jet of this source appears to be deflected and bent at about 0.5 mas distance from the radio core via gravitational lensing or the magnetosphere of a second massive black hole \citep{Britzen:2024egp, Britzen:2025bww}. This meandering jet structure may also explain the origin of the Amaterasu event despite being positioned at an angular offset of $2.5^\circ$. The value of $R_{\rm ext}\sim0.5$ pc presented in our analysis is large compared to the usual estimates for the BLR region $R_{\rm BLR}\sim L_{\rm disk, 45}^{0.5}$ cm \citep{Tavecchio:2008vq}. This may be interpreted as the emission region lying outside the BLR or at the edge, resulting in a low BLR photon density, as considered in our model.

A proton flux at the highest energy UHECR spectrum, although subdominant, cannot be ruled out and may originate from a distinct source class. Luminous AGNs or GRBs are probable source candidates for a proton primary since heavier nuclei are more susceptible to photodisintegration in such sources \citep{Unger:2015laa, Kachelriess:2017tvs, Das:2020nvx}. In such cases, a hard spectral index can be expected due to increased interaction in the vicinity of the source. However, we restrict ourselves to a prudent choice of $\alpha=2$ motivated by the first-order Fermi acceleration. 

A neural network classifier at the TA collaboration excludes photon as the primary particle at 99.986\% C.L \citep{TelescopeArray:2023sbd}. For both proton and iron primaries, the average propagation distance at 244 EeV is $\approx 30$ Mpc. A comoving distance of $\approx 590$ Mpc for the blazar source of this event can thus be reconciled with LIV effects in extragalactic propagation. The absence of other sources within the angular uncertainty region of $1.4^\circ$ strengthens the case for the blazar hypothesis. For a proton primary to be associated with the blazar, a strong EGMF strength is required. The uncertainty region due to GMF becomes larger for heavier elements, making it difficult to reconcile with the blazar source. Constraints on the number density of UHECR sources emitting heavy nuclei have been studied \citep{Kuznetsov:2023jfw}. 
Moreover, in order to reproduce the observed degree of isotropy of cosmic rays at $\sim$ EeV energies, the average magnetic fields in cosmic voids must be $\sim0.1$ nG \citep{Hackstein:2016pwa}. The value we obtain is an order of magnitude higher, corresponding to that in filaments.


The source is located above the IceCube horizon, allowing the observation of an upgoing $\nu_\mu$ track-like event, reducing the atmospheric background. The IceCube-Gen2 radio array \citep{IceCube:2019pna, IceCube-Gen2:2020qha}, with 8 times more instrument volume than the current capacity, will have five times more sensitivity, thus increasing the chance of neutrino detection from this source. During increased X-ray and $\gamma$-ray activity, if the kinetic power in protons increases by an order of magnitude, IceCube-Gen2 will be able to detect an event within a few years of observation. The effective area of the ARCA detector of KM3NeT \citep{KM3Net:2016zxf}, assuming an isotropic neutrino flux, is comparable to that of IceCube. Several other next-generation neutrino telescopes, such as TAMBO in the southern hemisphere \citep{Romero-Wolf:2020pzh} and satellite-based detector POEMMA \citep{POEMMA:2020ykm}, will increase the sensitivity of PeV neutrinos in the northern hemisphere.

The $\gamma$-ray luminosity of the source obtained in our leptohadronic model in the energy range 100 MeV to 100 GeV is $L_\gamma\approx1.1\times 10^{45}$ erg s$^{-1}$. The number of sources in the luminosity–redshift phase space similar to PKS 1717+177 is very small \citep[cf. Fig. 1 in][for the same energy range]{Das:2020hev}, making a significant diffuse UHECR flux excess above 10 EeV unlikely from a population of similar sources. It has been shown that resolved $\gamma$-ray sources are insufficient to account for the population of sources producing the highest-energy cosmic rays, and there must exist a population of UHECR sources that lack $\gamma$-ray emission or are unresolved by the current-generation $\gamma$-ray telescopes \citep{Partenheimer:2024fiq}. The number of detected UHECR events with 
$E \gtrsim 100$ EeV is currently too small for a statistically robust correlation study with the known blazar population. No significant anisotropy has been detected at these energies \citep{dOrfeuil:2014qgw, PierreAuger:2014yba}, making it difficult to identify potential sources. This can be attributed to the dominance of heavier nuclei and the resulting smearing of directionality in cosmic magnetic fields \citep{TelescopeArray:2024oux}. Although no event similar to Amaterasu with a plausible blazar counterpart has been identified, magnetic deflections and limited statistics hinder firm conclusions.

LIV effects become significant above $\sim 10$ EeV for our choice of $\delta_{\rm had, 0}=10^{-21}$. UHE protons at $\gtrsim 0.1$ EeV up to 10 EeV, escaping the blazar jet, can interact with the CMB and EBL photons to give rise to a line-of-sight resolved cosmogenic $\gamma$-ray spectrum \citep{Essey:2010er, Das:2019gtu}. It has a universal shape for a fully developed electromagnetic cascade, peaking at TeV energies for a given injection spectrum of UHE protons. Since the EGMF is considered to be rather strong, the collimation of the UHECR beam is difficult to maintain. Hence, for an appreciable cosmogenic $\gamma$-ray spectrum, most of the interactions must happen near the source. If CTA or LHAASO detects a steady multi-TeV $\gamma$-ray flux, this may indicate cosmogenic photons, further hinting toward UHECR acceleration.

\section{Conclusions\label{sec:conclusions}}

Our study presents a multiwavelength analysis of the blazar potentially linked to one of the highest energy UHECR events recorded using a modern state-of-the-art surface detector array. The findings weave together cosmic rays, neutrinos, and $\gamma$-rays into a unified multimessenger narrative. Our analysis also uncovers a novel probe of beyond Standard Model physics through the possible imprints of Lorentz invariance violation. We show that UHE protons from this source can be deflected by a few degrees, arriving at Earth for a sufficiently strong EGMF. The interaction of the proton spectrum within the jet may lead to a detectable signature in the soft X-ray and sub-TeV $\gamma$-ray spectrum. The neutrino flux obtained in the steady state has a low event rate, making it difficult to detect by the exposure of currently operating detectors. However, our analysis suggests that next-generation neutrino telescopes can constrain UHECR acceleration from this source within a few years of observation, particularly if an increased activity is observed in X-ray or $\gamma$-ray wave band. This source provides strong motivation for a dedicated multiwavelength and multimessenger campaign.

\software{GAMERA \citep{Hahn:2015hhw, Hahn_2022}, CRPropa 3.2 \citep{AlvesBatista:2016vpy, AlvesBatista:2022vem}
          }

\begin{acknowledgments}
Numerical computations in this work were carried out at the Raman Research Institute computing facility.
\end{acknowledgments}





\bibliography{sample631}{}

\begin{thebibliography}{}
\expandafter\ifx\csname natexlab\endcsname\relax\def\natexlab#1{#1}\fi
\providecommand{\url}[1]{\href{#1}{#1}}
\providecommand{\dodoi}[1]{doi:~\href{http://doi.org/#1}{\nolinkurl{#1}}}
\providecommand{\doeprint}[1]{\href{http://ascl.net/#1}{\nolinkurl{http://ascl.net/#1}}}
\providecommand{\doarXiv}[1]{\href{https://arxiv.org/abs/#1}{\nolinkurl{https://arxiv.org/abs/#1}}}

\bibitem[{{Abbasi} {et~al.}(2010)}]{TA_2010}
{Abbasi}, R.~U., {et~al.} 2010, \prl, 104, 161101,
  \dodoi{10.1103/PhysRevLett.104.161101}

\bibitem[{Aloisio {et~al.}(2012)Aloisio, Berezinsky, \&
  Gazizov}]{Aloisio:2012ba}
Aloisio, R., Berezinsky, V., \& Gazizov, A. 2012, Astropart. Phys., 39-40, 129,
  \dodoi{10.1016/j.astropartphys.2012.09.007}

\bibitem[{Alves~Batista {et~al.}(2016)Alves~Batista, Dundovic, Erdmann,
  Kampert, Kuempel, M\"uller, Sigl, van Vliet, Walz, \&
  Winchen}]{AlvesBatista:2016vpy}
Alves~Batista, R., Dundovic, A., Erdmann, M., {et~al.} 2016, JCAP, 05, 038,
  \dodoi{10.1088/1475-7516/2016/05/038}

\bibitem[{Alves~Batista {et~al.}(2022)}]{AlvesBatista:2022vem}
Alves~Batista, R., {et~al.} 2022, JCAP, 09, 035,
  \dodoi{10.1088/1475-7516/2022/09/035}

\bibitem[{Anchordoqui(2019)}]{Anchordoqui:2018qom}
Anchordoqui, L.~A. 2019, Phys. Rept., 801, 1,
  \dodoi{10.1016/j.physrep.2019.01.002}

\bibitem[{Ansoldi {et~al.}(2018)}]{MAGIC:2018sak}
Ansoldi, S., {et~al.} 2018, Astrophys. J. Lett., 863, L10,
  \dodoi{10.3847/2041-8213/aad083}

\bibitem[{Atoyan \& Dermer(2001)}]{Atoyan:2001ey}
Atoyan, A., \& Dermer, C.~D. 2001, Phys. Rev. Lett., 87, 221102,
  \dodoi{10.1103/PhysRevLett.87.221102}

\bibitem[{Barnacka {et~al.}(2014)Barnacka, Moderski, Behera, Brun, \&
  Wagner}]{Barnacka:2013oxa}
Barnacka, A., Moderski, R., Behera, B., Brun, P., \& Wagner, S. 2014, Astron.
  Astrophys., 567, A113, \dodoi{10.1051/0004-6361/201322205}

\bibitem[{{Becker} {et~al.}(1994){Becker}, {White}, \&
  {Helfand}}]{1994ASPC...61..165B}
{Becker}, R.~H., {White}, R.~L., \& {Helfand}, D.~J. 1994, in Astronomical
  Society of the Pacific Conference Series, Vol.~61, Astronomical Data Analysis
  Software and Systems III, ed. D.~R. {Crabtree}, R.~J. {Hanisch}, \&
  J.~{Barnes}, 165

\bibitem[{{Berezinskii} \& {Ginzburg}(1981)}]{1981MNRAS.194....3B}
{Berezinskii}, V.~S., \& {Ginzburg}, V.~L. 1981, \mnras, 194, 3,
  \dodoi{10.1093/mnras/194.1.3}

\bibitem[{{Bianchi} {et~al.}(2011){Bianchi}, {Herald}, {Efremova}, {Girardi},
  {Zabot}, {Marigo}, {Conti}, \& {Shiao}}]{2011Ap&SS.335..161B}
{Bianchi}, L., {Herald}, J., {Efremova}, B., {et~al.} 2011, \apss, 335, 161,
  \dodoi{10.1007/s10509-010-0581-x}

\bibitem[{Bird {et~al.}(1995)}]{HIRES:1994ijd}
Bird, D.~J., {et~al.} 1995, Astrophys. J., 441, 144, \dodoi{10.1086/175344}

\bibitem[{Boettcher {et~al.}(2013)Boettcher, Reimer, Sweeney, \&
  Prakash}]{Boettcher:2013wxa}
Boettcher, M., Reimer, A., Sweeney, K., \& Prakash, A. 2013, Astrophys. J.,
  768, 54, \dodoi{10.1088/0004-637X/768/1/54}

\bibitem[{{Boettcher} \& {Schlickeiser}(1997)}]{Boettcher_1997}
{Boettcher}, M., \& {Schlickeiser}, R. 1997, \aap, 325, 866,
  \dodoi{10.48550/arXiv.astro-ph/9703069}

\bibitem[{Bourriche \& Capel(2024)}]{Bourriche:2024bbe}
Bourriche, N., \& Capel, F. 2024.
\newblock \doarXiv{2406.16483}

\bibitem[{Britzen {et~al.}(2024)}]{Britzen:2024egp}
Britzen, S., {et~al.} 2024, Mon. Not. Roy. Astron. Soc., 535, 2742,
  \dodoi{10.1093/mnras/stae2373}

\bibitem[{Britzen {et~al.}(2025)}]{Britzen:2025bww}
---. 2025, Astron. Astrophys., 695, A103, \dodoi{10.1051/0004-6361/202452530}

\bibitem[{{Chodorowski} {et~al.}(1992){Chodorowski}, {Zdziarski}, \&
  {Sikora}}]{Chodorowski_92}
{Chodorowski}, M.~J., {Zdziarski}, A.~A., \& {Sikora}, M. 1992, \apj, 400, 181,
  \dodoi{10.1086/171984}

\bibitem[{Coleman \& Glashow(1997)}]{Coleman:1997xq}
Coleman, S.~R., \& Glashow, S.~L. 1997, Phys. Lett. B, 405, 249,
  \dodoi{10.1016/S0370-2693(97)00638-2}

\bibitem[{Das {et~al.}(2020)Das, Gupta, \& Razzaque}]{Das:2019gtu}
Das, S., Gupta, N., \& Razzaque, S. 2020, Astrophys. J., 889, 149,
  \dodoi{10.3847/1538-4357/ab6131}

\bibitem[{Das {et~al.}(2021{\natexlab{a}})Das, Gupta, \&
  Razzaque}]{Das:2020hev}
---. 2021{\natexlab{a}}, Astrophys. J., 910, 100,
  \dodoi{10.3847/1538-4357/abe4cd}

\bibitem[{Das {et~al.}(2022{\natexlab{a}})Das, Gupta, \&
  Razzaque}]{Das:2022nyp}
---. 2022{\natexlab{a}}, Astron. Astrophys., 668, A146,
  \dodoi{10.1051/0004-6361/202244653}

\bibitem[{Das {et~al.}(2021{\natexlab{b}})Das, Razzaque, \&
  Gupta}]{Das:2020nvx}
Das, S., Razzaque, S., \& Gupta, N. 2021{\natexlab{b}}, Eur. Phys. J. C, 81,
  59, \dodoi{10.1140/epjc/s10052-021-08885-4}

\bibitem[{Das {et~al.}(2022{\natexlab{b}})Das, Razzaque, \&
  Gupta}]{Das:2021cdf}
---. 2022{\natexlab{b}}, Astron. Astrophys., 658, L6,
  \dodoi{10.1051/0004-6361/202142123}

\bibitem[{Dermer {et~al.}(2009)Dermer, Razzaque, Finke, \&
  Atoyan}]{Dermer:2008cy}
Dermer, C.~D., Razzaque, S., Finke, J.~D., \& Atoyan, A. 2009, New J. Phys.,
  11, 065016, \dodoi{10.1088/1367-2630/11/6/065016}

\bibitem[{d'Orfeuil {et~al.}(2014)d'Orfeuil, Allard, Lachaud, Parizot,
  Blaksley, \& Nagataki}]{dOrfeuil:2014qgw}
d'Orfeuil, B.~R., Allard, D., Lachaud, C., {et~al.} 2014, Astron. Astrophys.,
  567, A81, \dodoi{10.1051/0004-6361/201423462}

\bibitem[{Ehlert {et~al.}(2024)Ehlert, van Vliet, Oikonomou, \&
  Winter}]{Ehlert:2023btz}
Ehlert, D., van Vliet, A., Oikonomou, F., \& Winter, W. 2024, JCAP, 02, 022,
  \dodoi{10.1088/1475-7516/2024/02/022}

\bibitem[{{Eichler}(1979)}]{1979ApJ...232..106E}
{Eichler}, D. 1979, \apj, 232, 106, \dodoi{10.1086/157269}

\bibitem[{Essey {et~al.}(2011)Essey, Kalashev, Kusenko, \&
  Beacom}]{Essey:2010er}
Essey, W., Kalashev, O., Kusenko, A., \& Beacom, J.~F. 2011, Astrophys. J.,
  731, 51, \dodoi{10.1088/0004-637X/731/1/51}

\bibitem[{Fargion {et~al.}(2024)Fargion, De~Sanctis~Lucentini, \&
  Khlopov}]{Fargion:2024ujt}
Fargion, D., De~Sanctis~Lucentini, P.~G., \& Khlopov, M.~Y. 2024, Universe, 10,
  323, \dodoi{10.3390/universe10080323}

\bibitem[{Farrar(2025)}]{Farrar:2024zsm}
Farrar, G.~R. 2025, Phys. Rev. Lett., 134, 081003,
  \dodoi{10.1103/PhysRevLett.134.081003}

\bibitem[{{Fermi-LAT Collaboration}(2020)}]{Fermi-LAT:2019yla}
{Fermi-LAT Collaboration}. 2020, Astrophys. J. Suppl., 247, 33,
  \dodoi{10.3847/1538-4365/ab6bcb}

\bibitem[{{Fermi-LAT Collaboration}(2023)}]{2023arXiv230712546B}
---. 2023, arXiv e-prints, arXiv:2307.12546, \dodoi{10.48550/arXiv.2307.12546}

\bibitem[{Finke(2019)}]{Finke:2018pkl}
Finke, J.~D. 2019, Astrophys. J., 870, 28, \dodoi{10.3847/1538-4357/aaf00c}

\bibitem[{Gilmore {et~al.}(2012)Gilmore, Somerville, Primack, \&
  Dominguez}]{Gilmore:2011ks}
Gilmore, R.~C., Somerville, R.~S., Primack, J.~R., \& Dominguez, A. 2012, Mon.
  Not. Roy. Astron. Soc., 422, 3189, \dodoi{10.1111/j.1365-2966.2012.20841.x}

\bibitem[{{Giommi} {et~al.}(2012){Giommi}, {Polenta}, {L{\"a}hteenm{\"a}ki},
  {Thompson}, {Capalbi}, {Cutini}, {Gasparrini}, {Gonz{\'a}lez-Nuevo},
  {Le{\'o}n-Tavares}, {L{\'o}pez-Caniego}, {Mazziotta}, {Monte}, {Perri},
  {Rain{\`o}}, {Tosti}, {Tramacere}, {Verrecchia}, {Aller}, {Aller},
  {Angelakis}, {Bastieri}, {Berdyugin}, {Bonaldi}, {Bonavera}, {Burigana},
  {Burrows}, {Buson}, {Cavazzuti}, {Chincarini}, {Colafrancesco}, {Costamante},
  {Cuttaia}, {D'Ammando}, {de Zotti}, {Frailis}, {Fuhrmann}, {Galeotta},
  {Gargano}, {Gehrels}, {Giglietto}, {Giordano}, {Giroletti}, {Keih{\"a}nen},
  {King}, {Krichbaum}, {Lasenby}, {Lavonen}, {Lawrence}, {Leto}, {Lindfors},
  {Mandolesi}, {Massardi}, {Max-Moerbeck}, {Michelson}, {Mingaliev}, {Natoli},
  {Nestoras}, {Nieppola}, {Nilsson}, {Partridge}, {Pavlidou}, {Pearson},
  {Procopio}, {Rachen}, {Readhead}, {Reeves}, {Reimer}, {Reinthal},
  {Ricciardi}, {Richards}, {Riquelme}, {Saarinen}, {Sajina}, {Sandri},
  {Savolainen}, {Sievers}, {Sillanp{\"a}{\"a}}, {Sotnikova}, {Stevenson},
  {Tagliaferri}, {Takalo}, {Tammi}, {Tavagnacco}, {Terenzi}, {Toffolatti},
  {Tornikoski}, {Trigilio}, {Turunen}, {Umana}, {Ungerechts}, {Villa}, {Wu},
  {Zacchei}, {Zensus}, \& {Zhou}}]{2012A&A...541A.160G}
{Giommi}, P., {Polenta}, G., {L{\"a}hteenm{\"a}ki}, A., {et~al.} 2012, \aap,
  541, A160, \dodoi{10.1051/0004-6361/201117825}

\bibitem[{Globus {et~al.}(2008)Globus, Allard, \& Parizot}]{Globus:2007bi}
Globus, N., Allard, D., \& Parizot, E. 2008, Astron. Astrophys., 479, 97,
  \dodoi{10.1051/0004-6361:20078653}

\bibitem[{Gould \& Schr\'eder(1967)}]{Gould_1967}
Gould, R.~J., \& Schr\'eder, G.~P. 1967, Phys. Rev., 155, 1404,
  \dodoi{10.1103/PhysRev.155.1404}

\bibitem[{{Greisen}(1966)}]{Greisen_1966}
{Greisen}, K. 1966, \prl, 16, 748, \dodoi{10.1103/PhysRevLett.16.748}

\bibitem[{Gueta(2021)}]{Gueta:2021vO}
Gueta, O. 2021, in Proceedings of 37th International Cosmic Ray Conference
  {\textemdash} PoS(ICRC2021), Vol. 395, 885, \dodoi{10.22323/1.395.0885}

\bibitem[{Hackstein {et~al.}(2016)Hackstein, Vazza, Br\"uggen, Sigl, \&
  Dundovic}]{Hackstein:2016pwa}
Hackstein, S., Vazza, F., Br\"uggen, M., Sigl, G., \& Dundovic, A. 2016, Mon.
  Not. Roy. Astron. Soc., 462, 3660, \dodoi{10.1093/mnras/stw1903}

\bibitem[{Hahn(2016)}]{Hahn:2015hhw}
Hahn, J. 2016, PoS, ICRC2015, 917, \dodoi{10.22323/1.236.0917}

\bibitem[{{Hahn} {et~al.}(2022){Hahn}, {Romoli}, \& {Breuhaus}}]{Hahn_2022}
{Hahn}, J., {Romoli}, C., \& {Breuhaus}, M. 2022, {GAMERA: Source modeling in
  gamma astronomy}, Astrophysics Source Code Library, record ascl:2203.007

\bibitem[{Harari {et~al.}(2014)Harari, Mollerach, \& Roulet}]{Harari:2013pea}
Harari, D., Mollerach, S., \& Roulet, E. 2014, Phys. Rev. D, 89, 123001,
  \dodoi{10.1103/PhysRevD.89.123001}

\bibitem[{{Healey} {et~al.}(2007){Healey}, {Romani}, {Taylor}, {Sadler},
  {Ricci}, {Murphy}, {Ulvestad}, \& {Winn}}]{2007ApJS..171...61H}
{Healey}, S.~E., {Romani}, R.~W., {Taylor}, G.~B., {et~al.} 2007, \apjs, 171,
  61, \dodoi{10.1086/513742}

\bibitem[{{IceCube Collaboration}(2018)}]{IceCube:2018cha}
{IceCube Collaboration}. 2018, Science, 361, 147,
  \dodoi{10.1126/science.aat2890}

\bibitem[{{IceCube Collaboration}(2019{\natexlab{a}})}]{IceCube:2018ndw}
---. 2019{\natexlab{a}}, Eur. Phys. J. C, 79, 234,
  \dodoi{10.1140/epjc/s10052-019-6680-0}

\bibitem[{{IceCube Collaboration}(2019{\natexlab{b}})}]{IceCube:2019pna}
---. 2019{\natexlab{b}}.
\newblock \doarXiv{1911.02561}

\bibitem[{{IceCube Collaboration}(2021)}]{IceCube-Gen2:2020qha}
---. 2021, J. Phys. G, 48, 060501, \dodoi{10.1088/1361-6471/abbd48}

\bibitem[{{IceCube Collaboration} {et~al.}(2018){IceCube Collaboration},
  {Fermi-LAT Collaboration}, {MAGIC Collaboration}, {et~al.}}]{IceCube:2018dnn}
{IceCube Collaboration}, {Fermi-LAT Collaboration}, {MAGIC Collaboration},
  {et~al.} 2018, Science, 361, eaat1378, \dodoi{10.1126/science.aat1378}

\bibitem[{Jansson \& Farrar(2012)}]{Jansson:2012pc}
Jansson, R., \& Farrar, G.~R. 2012, Astrophys. J., 757, 14,
  \dodoi{10.1088/0004-637X/757/1/14}

\bibitem[{Jiang {et~al.}(2025)Jiang, Liao, Rui, Fan, \& Wei}]{Jiang:2025hqb}
Jiang, X., Liao, N.-H., Rui, X., Fan, Y.-Z., \& Wei, D.-M. 2025.
\newblock \doarXiv{2505.04160}

\bibitem[{Kachelrie\ss{} {et~al.}(2017)Kachelrie\ss{}, Kalashev, Ostapchenko,
  \& Semikoz}]{Kachelriess:2017tvs}
Kachelrie\ss{}, M., Kalashev, O., Ostapchenko, S., \& Semikoz, D.~V. 2017,
  Phys. Rev. D, 96, 083006, \dodoi{10.1103/PhysRevD.96.083006}

\bibitem[{Kawamuro {et~al.}(2018)}]{Kawamuro:2018eky}
Kawamuro, T., {et~al.} 2018, Astrophys. J. Suppl., 238, 32,
  \dodoi{10.3847/1538-4365/aad1ef}

\bibitem[{Keivani {et~al.}(2018)}]{Keivani:2018rnh}
Keivani, A., {et~al.} 2018, Astrophys. J., 864, 84,
  \dodoi{10.3847/1538-4357/aad59a}

\bibitem[{Kelner \& Aharonian(2008)}]{Kelner:2008ke}
Kelner, S.~R., \& Aharonian, F.~A. 2008, Phys. Rev. D, 78, 034013,
  \dodoi{10.1103/PhysRevD.82.099901}

\bibitem[{Kim {et~al.}(2023)Kim, Ivanov, Jui, \& Thomson}]{Kim:2023eul}
Kim, J., Ivanov, D., Jui, C., \& Thomson, G. 2023, EPJ Web Conf., 283, 02005,
  \dodoi{10.1051/epjconf/202328302005}

\bibitem[{{KM3NeT Collaboration}(2016)}]{KM3Net:2016zxf}
{KM3NeT Collaboration}. 2016, J. Phys. G, 43, 084001,
  \dodoi{10.1088/0954-3899/43/8/084001}

\bibitem[{{Kotera} \& {Olinto}(2011)}]{Kotera_2011}
{Kotera}, K., \& {Olinto}, A.~V. 2011, \araa, 49, 119,
  \dodoi{10.1146/annurev-astro-081710-102620}

\bibitem[{Kuznetsov(2024)}]{Kuznetsov:2023jfw}
Kuznetsov, M.~Y. 2024, JCAP, 04, 042, \dodoi{10.1088/1475-7516/2024/04/042}

\bibitem[{Lang(2024)}]{Lang:2024jmc}
Lang, R.~G. 2024, JCAP, 11, 023, \dodoi{10.1088/1475-7516/2024/11/023}

\bibitem[{{Mannheim} {et~al.}(1992){Mannheim}, {Stanev}, \&
  {Biermann}}]{1992A&A...260L...1M}
{Mannheim}, K., {Stanev}, T., \& {Biermann}, P.~L. 1992, \aap, 260, L1

\bibitem[{{M{\"u}cke} {et~al.}(2000){M{\"u}cke}, {Engel}, {Rachen},
  {Protheroe}, \& {Stanev}}]{Mucke_00}
{M{\"u}cke}, A., {Engel}, R., {Rachen}, J.~P., {Protheroe}, R.~J., \& {Stanev},
  T. 2000, Computer Physics Communications, 124, 290,
  \dodoi{10.1016/S0010-4655(99)00446-4}

\bibitem[{Murase(2012)}]{Murase:2011yw}
Murase, K. 2012, Astrophys. J. Lett., 745, L16,
  \dodoi{10.1088/2041-8205/745/2/L16}

\bibitem[{Murase {et~al.}(2014)Murase, Inoue, \& Dermer}]{Murase:2014foa}
Murase, K., Inoue, Y., \& Dermer, C.~D. 2014, Phys. Rev. D, 90, 023007,
  \dodoi{10.1103/PhysRevD.90.023007}

\bibitem[{Murase {et~al.}(2025)Murase, Narita, \& Yin}]{Murase:2025uwv}
Murase, K., Narita, Y., \& Yin, W. 2025.
\newblock \doarXiv{2504.15272}

\bibitem[{Murase \& Stecker(2023)}]{Murase:2022feu}
Murase, K., \& Stecker, F.~W. 2023, 483, \dodoi{10.1142/9789811282645_0010}

\bibitem[{Muzio {et~al.}(2022)Muzio, Farrar, \& Unger}]{Muzio:2021zud}
Muzio, M.~S., Farrar, G.~R., \& Unger, M. 2022, Phys. Rev. D, 105, 023022,
  \dodoi{10.1103/PhysRevD.105.023022}

\bibitem[{Muzio {et~al.}(2019)Muzio, Unger, \& Farrar}]{Muzio:2019leu}
Muzio, M.~S., Unger, M., \& Farrar, G.~R. 2019, Phys. Rev. D, 100, 103008,
  \dodoi{10.1103/PhysRevD.100.103008}

\bibitem[{{Nieppola} {et~al.}(2007){Nieppola}, {Tornikoski},
  {L{\"a}hteenm{\"a}ki}, {Valtaoja}, {Hakala}, {Hovatta}, {Kotiranta},
  {Nummila}, {Ojala}, {Parviainen}, {Ranta}, {Saloranta}, {Torniainen}, \&
  {Tr{\"o}ller}}]{2007AJ....133.1947N}
{Nieppola}, E., {Tornikoski}, M., {L{\"a}hteenm{\"a}ki}, A., {et~al.} 2007,
  \aj, 133, 1947, \dodoi{10.1086/512609}

\bibitem[{Olinto {et~al.}(2021)}]{POEMMA:2020ykm}
Olinto, A.~V., {et~al.} 2021, JCAP, 06, 007,
  \dodoi{10.1088/1475-7516/2021/06/007}

\bibitem[{Padovani {et~al.}(2019)Padovani, Oikonomou, Petropoulou, Giommi, \&
  Resconi}]{Padovani:2019xcv}
Padovani, P., Oikonomou, F., Petropoulou, M., Giommi, P., \& Resconi, E. 2019,
  Mon. Not. Roy. Astron. Soc., 484, L104, \dodoi{10.1093/mnrasl/slz011}

\bibitem[{Partenheimer {et~al.}(2024)Partenheimer, Fang, Alves~Batista, \&
  Menezes~de Almeida}]{Partenheimer:2024fiq}
Partenheimer, A., Fang, K., Alves~Batista, R., \& Menezes~de Almeida, R. 2024,
  Astrophys. J. Lett., 967, L15, \dodoi{10.3847/2041-8213/ad4359}

\bibitem[{{Pierre Auger
  Collaboration}(2015{\natexlab{a}})}]{PierreAuger:2015eyc}
{Pierre Auger Collaboration}. 2015{\natexlab{a}}, Nucl. Instrum. Meth. A, 798,
  172, \dodoi{10.1016/j.nima.2015.06.058}

\bibitem[{{Pierre Auger
  Collaboration}(2015{\natexlab{b}})}]{PierreAuger:2014yba}
---. 2015{\natexlab{b}}, Astrophys. J., 804, 15,
  \dodoi{10.1088/0004-637X/804/1/15}

\bibitem[{{Pierre Auger Collaboration}(2017)}]{PierreAuger:2016use}
---. 2017, JCAP, 04, 038, \dodoi{10.1088/1475-7516/2017/04/038}

\bibitem[{{Pierre Auger Collaboration}(2022)}]{PierreAuger:2021tog}
---. 2022, JCAP, 01, 023, \dodoi{10.1088/1475-7516/2022/01/023}

\bibitem[{{Planck Collaboration} {et~al.}(2014){Planck Collaboration}, {Ade},
  {Aghanim}, {Arg{\"u}eso}, {Armitage-Caplan}, {Arnaud}, {Ashdown},
  {Atrio-Barandela}, {Aumont}, {Baccigalupi}, {Banday}, {Barreiro}, {Bartlett},
  {Battaner}, {Beelen}, {Benabed}, {Beno{\^\i}t}, {Benoit-L{\'e}vy}, {Bernard},
  {Bersanelli}, {Bielewicz}, {Bobin}, {Bock}, {Bonaldi}, {Bonavera}, {Bond},
  {Borrill}, {Bouchet}, {Bridges}, {Bucher}, {Burigana}, {Butler}, {Cardoso},
  {Carvalho}, {Catalano}, {Challinor}, {Chamballu}, {Chen}, {Chiang}, {Chiang},
  {Christensen}, {Church}, {Clemens}, {Clements}, {Colombi}, {Colombo},
  {Couchot}, {Coulais}, {Crill}, {Curto}, {Cuttaia}, {Danese}, {Davies},
  {Davis}, {de Bernardis}, {de Rosa}, {de Zotti}, {Delabrouille}, {Delouis},
  {D{\'e}sert}, {Dickinson}, {Diego}, {Dole}, {Donzelli}, {Dor{\'e}},
  {Douspis}, {Dupac}, {Efstathiou}, {En{\ss}lin}, {Eriksen}, {Finelli},
  {Forni}, {Frailis}, {Franceschi}, {Galeotta}, {Ganga}, {Giard}, {Giardino},
  {Giraud-H{\'e}raud}, {Gonz{\'a}lez-Nuevo}, {G{\'o}rski}, {Gratton},
  {Gregorio}, {Gruppuso}, {Hansen}, {Hanson}, {Harrison},
  {Henrot-Versill{\'e}}, {Hern{\'a}ndez-Monteagudo}, {Herranz}, {Hildebrandt},
  {Hivon}, {Hobson}, {Holmes}, {Hornstrup}, {Hovest}, {Huffenberger}, {Jaffe},
  {Jaffe}, {Jones}, {Juvela}, {Keih{\"a}nen}, {Keskitalo}, {Kisner}, {Kneissl},
  {Knoche}, {Knox}, {Kunz}, {Kurki-Suonio}, {Lagache}, {L{\"a}hteenm{\"a}ki},
  {Lamarre}, {Lasenby}, {Laureijs}, {Lawrence}, {Leahy}, {Leonardi},
  {Le{\'o}n-Tavares}, {Leroy}, {Lesgourgues}, {Liguori}, {Lilje},
  {Linden-V{\o}rnle}, {L{\'o}pez-Caniego}, {Lubin}, {Mac{\'\i}as-P{\'e}rez},
  {Maffei}, {Maino}, {Mandolesi}, {Maris}, {Marshall}, {Martin},
  {Mart{\'\i}nez-Gonz{\'a}lez}, {Masi}, {Massardi}, {Matarrese}, {Matthai},
  {Mazzotta}, {McGehee}, {Meinhold}, {Melchiorri}, {Mendes}, {Mennella},
  {Migliaccio}, {Mitra}, {Miville-Desch{\^e}nes}, {Moneti}, {Montier},
  {Morgante}, {Mortlock}, {Munshi}, {Murphy}, {Naselsky}, {Nati}, {Natoli},
  {Negrello}, {Netterfield}, {N{\o}rgaard-Nielsen}, {Noviello}, {Novikov},
  {Novikov}, {O'Dwyer}, {Osborne}, {Oxborrow}, {Paci}, {Pagano}, {Pajot},
  {Paladini}, {Paoletti}, {Partridge}, {Pasian}, {Patanchon}, {Pearson},
  {Perdereau}, {Perotto}, {Perrotta}, {Piacentini}, {Piat}, {Pierpaoli},
  {Pietrobon}, {Plaszczynski}, {Pointecouteau}, {Polenta}, {Ponthieu}, {Popa},
  {Poutanen}, {Pratt}, {Pr{\'e}zeau}, {Prunet}, {Puget}, {Rachen}, {Reach},
  {Rebolo}, {Reinecke}, {Remazeilles}, {Renault}, {Ricciardi}, {Riller},
  {Ristorcelli}, \& {Rocha}}]{2014A&A...571A..28P}
{Planck Collaboration}, {Ade}, P.~A.~R., {Aghanim}, N., {et~al.} 2014, \aap,
  571, A28, \dodoi{10.1051/0004-6361/201321524}

\bibitem[{Prince {et~al.}(2024)Prince, Das, Gupta, Majumdar, \&
  Czerny}]{Prince:2023bhj}
Prince, R., Das, S., Gupta, N., Majumdar, P., \& Czerny, B. 2024, Mon. Not.
  Roy. Astron. Soc., 527, 8746, \dodoi{10.1093/mnras/stad3804}

\bibitem[{Pshirkov {et~al.}(2011)Pshirkov, Tinyakov, Kronberg, \&
  Newton-McGee}]{Pshirkov:2011um}
Pshirkov, M.~S., Tinyakov, P.~G., Kronberg, P.~P., \& Newton-McGee, K.~J. 2011,
  Astrophys. J., 738, 192, \dodoi{10.1088/0004-637X/738/2/192}

\bibitem[{{Pushkarev} {et~al.}(2009){Pushkarev}, {Kovalev}, {Lister}, \&
  {Savolainen}}]{Pushkarev_2009}
{Pushkarev}, A.~B., {Kovalev}, Y.~Y., {Lister}, M.~L., \& {Savolainen}, T.
  2009, \aap, 507, L33, \dodoi{10.1051/0004-6361/200913422}

\bibitem[{Riehn {et~al.}(2020)Riehn, Engel, Fedynitch, Gaisser, \&
  Stanev}]{Riehn:2019jet}
Riehn, F., Engel, R., Fedynitch, A., Gaisser, T.~K., \& Stanev, T. 2020, Phys.
  Rev. D, 102, 063002, \dodoi{10.1103/PhysRevD.102.063002}

\bibitem[{Romero-Wolf {et~al.}(2020)}]{Romero-Wolf:2020pzh}
Romero-Wolf, A., {et~al.} 2020, in {Latin American Strategy Forum for Research
  Infrastructure}.
\newblock \doarXiv{2002.06475}

\bibitem[{Sarmah {et~al.}(2025)Sarmah, Das, Borah, Chakraborty, \&
  Mehta}]{Sarmah:2024ffy}
Sarmah, P., Das, N., Borah, D., Chakraborty, S., \& Mehta, P. 2025, Phys. Rev.
  D, 111, 083048, \dodoi{10.1103/PhysRevD.111.083048}

\bibitem[{Sigl {et~al.}(2004)Sigl, Miniati, \& Ensslin}]{Sigl:2004yk}
Sigl, G., Miniati, F., \& Ensslin, T.~A. 2004, Phys. Rev. D, 70, 043007,
  \dodoi{10.1103/PhysRevD.70.043007}

\bibitem[{{Sikora} {et~al.}(1987){Sikora}, {Kirk}, {Begelman}, \&
  {Schneider}}]{Sikora_1987}
{Sikora}, M., {Kirk}, J.~G., {Begelman}, M.~C., \& {Schneider}, P. 1987, \apjl,
  320, L81, \dodoi{10.1086/184980}

\bibitem[{Stecker(1968)}]{Stecker:1968uc}
Stecker, F.~W. 1968, Phys. Rev. Lett., 21, 1016,
  \dodoi{10.1103/PhysRevLett.21.1016}

\bibitem[{{Stecker} {et~al.}(1991){Stecker}, {Done}, {Salamon}, \&
  {Sommers}}]{1991PhRvL..66.2697S}
{Stecker}, F.~W., {Done}, C., {Salamon}, M.~H., \& {Sommers}, P. 1991, \prl,
  66, 2697, \dodoi{10.1103/PhysRevLett.66.2697}

\bibitem[{Stettner(2020)}]{Stettner:2019tok}
Stettner, J. 2020, PoS, ICRC2019, 1017, \dodoi{10.22323/1.358.1017}

\bibitem[{{Stratta} {et~al.}(2011){Stratta}, {Capalbi}, {Giommi}, {Primavera},
  {Cutini}, \& {Gasparrini}}]{Stratta_2011}
{Stratta}, G., {Capalbi}, M., {Giommi}, P., {et~al.} 2011, arXiv e-prints,
  arXiv:1103.0749, \dodoi{10.48550/arXiv.1103.0749}

\bibitem[{Szabo \& Protheroe(1994)}]{Szabo:1994qx}
Szabo, A.~P., \& Protheroe, R.~J. 1994, Astropart. Phys., 2, 375,
  \dodoi{10.1016/0927-6505(94)90027-2}

\bibitem[{Tavecchio \& Ghisellini(2008)}]{Tavecchio:2008vq}
Tavecchio, F., \& Ghisellini, G. 2008, Mon. Not. Roy. Astron. Soc., 386, 945,
  \dodoi{10.1111/j.1365-2966.2008.13072.x}

\bibitem[{{Tavecchio} {et~al.}(2010){Tavecchio}, {Ghisellini}, {Ghirlanda},
  {Foschini}, \& {Maraschi}}]{2010MNRAS.401.1570T}
{Tavecchio}, F., {Ghisellini}, G., {Ghirlanda}, G., {Foschini}, L., \&
  {Maraschi}, L. 2010, \mnras, 401, 1570,
  \dodoi{10.1111/j.1365-2966.2009.15784.x}

\bibitem[{{Telescope Array Collaboration}(2016)}]{TelescopeArray:2015dcv}
{Telescope Array Collaboration}. 2016, Astropart. Phys., 80, 131,
  \dodoi{10.1016/j.astropartphys.2016.04.002}

\bibitem[{{Telescope Array Collaboration}(2023)}]{TelescopeArray:2023sbd}
---. 2023, Science, 382, abo5095, \dodoi{10.1126/science.abo5095}

\bibitem[{{Telescope Array Collaboration}(2024)}]{TelescopeArray:2024oux}
---. 2024, Phys. Rev. Lett., 133, 041001,
  \dodoi{10.1103/PhysRevLett.133.041001}

\bibitem[{Tully {et~al.}(2008)Tully, Shaya, Karachentsev, Courtois, Kocevski,
  Rizzi, \& Peel}]{Tully:2007ue}
Tully, R.~B., Shaya, E.~J., Karachentsev, I.~D., {et~al.} 2008, Astrophys. J.,
  676, 184, \dodoi{10.1086/527428}

\bibitem[{Unger \& Farrar(2024{\natexlab{a}})}]{Unger:2023hnu}
Unger, M., \& Farrar, G.~R. 2024{\natexlab{a}}, Astrophys. J. Lett., 962, L5,
  \dodoi{10.3847/2041-8213/ad1ced}

\bibitem[{Unger \& Farrar(2024{\natexlab{b}})}]{Unger:2023lob}
---. 2024{\natexlab{b}}, Astrophys. J., 970, 95,
  \dodoi{10.3847/1538-4357/ad4a54}

\bibitem[{Unger {et~al.}(2015)Unger, Farrar, \& Anchordoqui}]{Unger:2015laa}
Unger, M., Farrar, G.~R., \& Anchordoqui, L.~A. 2015, Phys. Rev. D, 92, 123001,
  \dodoi{10.1103/PhysRevD.92.123001}

\bibitem[{Vazza {et~al.}(2014)Vazza, Br\"uggen, Gheller, \&
  Wang}]{Vazza:2014jga}
Vazza, F., Br\"uggen, M., Gheller, C., \& Wang, P. 2014, Mon. Not. Roy. Astron.
  Soc., 445, 3706, \dodoi{10.1093/mnras/stu1896}

\bibitem[{Vernetto(2016)}]{Vernetto:2016gro}
Vernetto, S. 2016, J. Phys. Conf. Ser., 718, 052043,
  \dodoi{10.1088/1742-6596/718/5/052043}

\bibitem[{{Wright} \& {Otrupcek}(1992)}]{1992BICDS..41...47W}
{Wright}, A., \& {Otrupcek}, R. 1992, Bulletin d'Information du Centre de
  Donnees Stellaires, 41, 47

\bibitem[{{Wright} {et~al.}(2010){Wright}, {Eisenhardt}, {Mainzer}, {Ressler},
  {Cutri}, {Jarrett}, {Kirkpatrick}, {Padgett}, {McMillan}, {Skrutskie},
  {Stanford}, {Cohen}, {Walker}, {Mather}, {Leisawitz}, {Gautier}, {McLean},
  {Benford}, {Lonsdale}, {Blain}, {Mendez}, {Irace}, {Duval}, {Liu}, {Royer},
  {Heinrichsen}, {Howard}, {Shannon}, {Kendall}, {Walsh}, {Larsen}, {Cardon},
  {Schick}, {Schwalm}, {Abid}, {Fabinsky}, {Naes}, \&
  {Tsai}}]{2010AJ....140.1868W}
{Wright}, E.~L., {Eisenhardt}, P. R.~M., {Mainzer}, A.~K., {et~al.} 2010, \aj,
  140, 1868, \dodoi{10.1088/0004-6256/140/6/1868}

\bibitem[{{Zatsepin} \& {Kuz'min}(1966)}]{Zatsepin_1966}
{Zatsepin}, G.~T., \& {Kuz'min}, V.~A. 1966, Soviet Journal of Experimental and
  Theoretical Physics Letters, 4, 78

\bibitem[{Zhang {et~al.}(2024)Zhang, Murase, Ekanger, Bhattacharya, \&
  Horiuchi}]{Zhang:2024sjp}
Zhang, B.~T., Murase, K., Ekanger, N., Bhattacharya, M., \& Horiuchi, S. 2024.
\newblock \doarXiv{2405.17409}

\end{thebibliography}
\bibliographystyle{aasjournal}



\end{document}